\documentstyle[twoside,fleqn,espcrc2]{article}
\def\beq{\begin{equation}}
\def\eeq{\end{equation}}
\def\bea{\begin{array}}
\def\eea{\end{array}}
\def\beqa{\begin{eqnarray}}
\def\eeqa{\end{eqnarray}}

\def\u1{{U(1)}}
\def\su2{{SU(2)}}
\newcommand{\re}{\relax{\rm I\kern-.18em R}}

\newcommand{\AmS}{{\protect\the\textfont2
  A\kern-.1667emt\lower.5ex\hbox{M}\kern-.125emS}}

\title{Strongly coupled 't Hooft model on the lattice
\thanks{Presented by F. Berruto at 
Lattice '99, Pisa, Italy.}
\vskip-3cm\hfill\small DFUPG-65-99\vskip2.6cm
}
\author{F. Berruto, E. Coletti, G. Grignani and P. Sodano
\vspace{6pt}\\ {Dipartimento di Fisica and Sezione I.N.F.N.,
\vskip0.1cm Universit\'a di Perugia, Via Pascoli I-06123 Perugia, Italy}}
\begin{document}
\begin{abstract}
A lattice strong coupling calculation of the spectrum and chiral condensate 
of the 't Hooft model is presented. The agreement with the results of the continuum 
theory is strikingly good even at the fourth order in the strong coupling expansions.
\end{abstract}
\maketitle
\section{Introduction}
Several attempts have been made to analyze the spectrum of two-dimensional 
quantum chromodynamics (QCD$_2$). 
't Hooft~\cite{q2} was able to find the meson spectrum 
for $U({\cal N}_c)$ QCD$_2$ using the $1/{\cal N}_c$ expansion. The results,
obtained in the weak coupling limit with $g^2{\cal N}_c$ fixed and ${\cal N}_c$ approaching 
infinity, revealed that there is a spectrum of color singlet mesonic 
states with equal energy spacing. Later, Witten~\cite{q3} explained how 
to fit baryons into this picture, showing that they can be interpreted 
as the 't Hooft-Polyakov monopoles of the theory.

In this contribution we report on a lattice computation on the spectrum 
and chiral condensate of the 't Hooft model in the strong coupling limit. The strong coupling limit 
of gauge theories is highly nonuniversal; in spite of this difficulty there exist strong coupling 
computations which claim some degree of success~\cite{q32}. 
We compute analytically the mass of the scalar and the 
pseudoscalar mesons and the chiral condensate up to the fourth 
order in a strong coupling expansion.
We find also that Witten's interpretation for the 
baryons is consistent with our lattice results.

The continuum one-flavor 't Hooft model is defined by the Euclidean action
\begin{equation}
S=\int \left[\overline{\psi}_{a}\gamma_{\mu}(\partial_{\mu}
\psi_{a}+A_{a\mu}^{b}
\psi_{b})-
{1\over 4g^{2}}F_{\mu \nu a}^{b}F_{\mu \nu b}^{a}\right]d^{2}x
\label{tho}
\end{equation}
where $a,b=1,\ldots {\cal N}_{c}$.

The Hamiltonian and the Gauss constraints are 
\begin{eqnarray}
H=\int \left[ {g^{2}\over 2}(E^A(x))^2+
\overline{\psi}_{a}\alpha(i\partial_x\psi_{a}+A_{a\ x}^{b}\psi_{b})\right]dx 
\label{hth}\\
\partial_x E^A(x)+ig[A^A(x),E^A(x)]+\psi^\dagger_a T^A_{ab}\psi_b(x)\sim0
\label{gaussc}
\end{eqnarray}
with $A=0,\dots,{\cal N}_c^2-1$ and  
the electric field operators $E^A(x)$ satisfying the group algebra.
The lattice Hamiltonian and Gauss constraints reproducing Eqs.(\ref{hth},
\ref{gaussc}) in the naive 
continuum limit, read
\begin{eqnarray}
H={g^2a\over 2}\sum_{x=1}^{N}E^{A}_xE^{A}_x-{it\over 2a}
\left[ R -L \right]\quad ,
\label{lhth} \\
E^A_x-U^\dagger(x-1)E^A_{x-1}U(x-1)+\cr
+\sum_{a,b=1}^{{\cal N}_c}\psi^\dagger_a T^A_{ab}\psi_b(x)-
\frac{{\cal N}_c}{2}\delta^{A,0}\sim0
\label{gaussl}
\end{eqnarray}
where the right and left hopping operators are defined ($L=R^{\dagger}$) by
\begin{equation}
R=\sum_{x=1}^{N}R_{x}= 
\sum_{x=1}^{N} \sum_{a,b=1}^{{\cal N}_c}
\psi_{a,x+1}^{\dag}U_{ab}(x)\psi_{b,x}\quad 
\label{rr}
\end{equation}
and the matrix $U(x)$, associated with the link $[x,x+1]$, 
is a group element of $U({\cal N}_c)$ in the fundamental 
representation.

The Hamiltonian (\ref{lhth}), rescaled by the factor $g^{2}a/ 2$, 
can be written as
\begin{equation}
H=H_{0}+\epsilon H_{h} 
\label{lbhth}
\end{equation}
with $H_{0}=\sum_{x=1}^{N}E_{x}^{A}E_{x}^{A}$, $H_{h}=-i(R-L)$ and 
$\epsilon=t/g^{2}a^{2}$ the expansion parameter.
Since $H_{0}$ and $H_{h}$ are both gauge invariant, if one finds a 
gauge invariant 
eigenstate of $H_{0}$, perturbations in $H_{h}$ retain gauge 
invariance. Due to the Gauss law constraints (\ref{gaussl}) the lowest energy eigenstate 
of $H_{0}$ is a color singlet with a density of states of  
${\cal N}_{c}/2$ fermions per site. A color singlet at 
a given site can be formed either by leaving it unoccupied or by 
putting on it ${\cal N}_{c}$ fermions by means of the creation operator
$
S_{+}(x)=\epsilon_{a_{1}\ldots a_{{\cal N}_{c}}}
\psi_{a_{1} x}^{\dagger}\ldots \psi_{a_{{\cal N}_{c}}}^{\dagger}\quad .
$
The $N/2$ singlets can be distributed arbitrarily 
among the $N$ sites so that there are
$N!/(N/2)!$ degenerate ground states. 
The local fermion number operator
\begin{equation}
\rho_{x}=\sum_{a=1}^{{\cal N}_{c}}\psi_{a x}^{\dagger}
\psi_{a x}-{\cal N}_{c}/2
\end{equation}
takes the value $+{\cal N}_{c}/2$ on occupied sites and 
$-{\cal N}_{c}/2$ on empty sites.

\section{Hadron spectrum}

First order perturbations to the vacuum energy vanish. 
The ground state degeneracy is removed at the second order in 
the strong coupling expansion. 
The vacuum energy - at order $\epsilon^2$ - reads
\begin{equation}
E_{0}^{(2)}=\langle H_{h}^{\dagger}\frac{\Pi}{E_{0}^{(0)}-H_{0}}H_{h} \rangle\quad ,
\label{soen}
\end{equation}
where the expectation values are defined on the degenerate subspace and 
$\Pi$ is the operator projecting on a set orthogonal to the degenerate ground states.
The commutator 
$
\left[H_{0},H_{h}\right]=C_{2}^{f}({\cal N}_{c})H_{h}
$,
where $C_{2}^{f}({\cal N}_{c})={\cal N}_{c}/2$ is 
the quadratic Casimir of the fundamental representation of $U({\cal N}_{c})$,
holds on the degenerate subspace.  
Using the commutator between $H_{0}$ and $H_h$  
from Eq.(\ref{soen}), one gets
\begin{equation}
E_{0}^{(2)}=-\frac{2}{C_{2}^{f}({\cal N}_{c})}\langle RL \rangle\ .
\label{seon2}
\end{equation}
The vacuum expectation value $\langle \ ,\ \rangle$ is the inner 
product in the full Hilbert space and is defined as 
$\langle \ ,\ \rangle=\prod_{x}\int dU_{x} (\ ,\ )$, where $dU$ is the Haar 
measure on the gauge group manifold and $(\ ,\ )$ is the fermion Fock 
space inner product. 
Performing the integrals over the group elements in Eq.(\ref{seon2}) 
the combination $RL$ can be written 
as a spin-${\cal N}_c/2$ Ising Hamiltonian: in fact, 
taking into account that products of $L_{x}$ and $R_{y}$ at different points have 
vanishing expectation values, 
one can rewrite Eq.(\ref{seon2}) as 
\begin{equation}
E_{0}^{(2)}=\frac{4}{{\cal N}_{c}^2}\ \langle\ \sum_{x=1}^{N}\rho(x)\rho(x+1) 
-\frac{1}{4}{\cal N}_{c}^{2}N\  \rangle\quad .
\label{ising}
\end{equation}
The Hamiltonian in Eq.(\ref{ising}) is an antiferromagnetic Ising Hamiltonian 
in the space of pure fermion states where $\rho_{x}=\pm{\cal N}_{c}/2$ and 
$\sum_{x=1}^{N}\rho_x=0$. 
The Hamiltonian in Eq.(\ref{ising}) has two degenerate ground states 
characterized by a fermion distribution 
$\rho_{x}=\pm{\cal N}_{c}/2 (-1)^{x}$ .

Let us now investigate the one-flavor 't Hooft model meson spectrum in the strong 
coupling limit. We evaluate the ground state energy up to the 
fourth order in the strong coupling expansion
\begin{eqnarray}
E_{g.s.}&=&\frac{g^2a}{2}(E_{g.s.}^{(0)}+\epsilon^2E_{g.s.}^{(2)}+
\epsilon^4E_{g.s.}^{(4)})\nonumber\\
&=&\frac{g^2a}{2}(-2{\cal N}_{c}^2N\epsilon^2+16{\cal N}_{c}^2N\epsilon^4)\ .
\label{gse}
\end{eqnarray}

The lowest lying excitations are a pseudoscalar and a scalar
created by the Fourier transform of the conserved gauge invariant currents 
at zero momentum $\sum_x j_1(x)=R+L$ and $\sum_x j_5(x)=R-L$, 
respectively. 
\begin{equation}
|P\rangle=(R+L)|g.s.\rangle\label{pe}\ ,\ \ |S\rangle=(R-L)|g.s.\rangle\label{se}
\end{equation}
At the zero-th order they are degenerate, but the degeneracy is removed 
at the second order in the strong coupling expansion.
The mass of these low lying excitations can be obtained by computing their
energies  and by subtracting the ground state energy (\ref{gse}). 

Up to the fourth order in $\epsilon$
the mass of the state $|P\rangle$ is given by
\begin{equation}
m_{P}=\frac{g^2a}{2}(\frac{1}{2}-4\epsilon^2+80\epsilon^4)
\label{mp}
\end{equation}
whereas the one of the scalar meson $|S\rangle$ is
\begin{equation}
m_{S}=\frac{g^2a}{2}(\frac{1}{2}+4\epsilon^2+80\epsilon^4).
\label{ms}
\end{equation}
In Eqs.(\ref{mp},\ref{ms}) the coupling constant has been rescaled according to 
$g^2\longrightarrow g^2{\cal N}_c$ so that
the strong coupling expansion parameter changes as  
$\epsilon\longrightarrow \frac{\epsilon}{{\cal N}_{c}}$. This rescaling 
is needed, since the meson masses are 
proportional to ${\cal N}_c$ at each order in the strong coupling expansion. 
This infinity  can be absorbed  in the definition of the coupling constant
to produce a smooth large-${\cal N}_c$ limit.

The spectrum exhibits also one baryon which can be created at zero momentum by acting 
on the ground state with the color singlet operator $B^\dagger$
\begin{equation}
|B>=B^\dagger|g.s.>=\sum_{x=1}^N\psi^\dagger_{1x}\psi^\dagger_{2x}\dots
\psi^\dagger_{{\cal N}_c x}|g.s.>\quad .
\end{equation}
At the zero-th order in the strong coupling expansion the baryon is massless, 
since the creation operator $B^\dagger$ does not contain any color flux ($H_0|B>=0$).
At the second order the baryon acquires a mass $m^{(2)}_B=g^2 a$. This result
is in agreement with Witten's conjecture~\cite{q3} 
that the baryons are the solitons of the theory:
baryons have a mass proportional to the inverse of the coupling constant (in 
strong coupling at the zero-th order they are massless) but acquire a mass proportional to 
$g^2$ already at the second order.

\section{Chiral symmetry breaking}

In the continuum 't Hooft model, the chiral symmetry is 
dynamically broken by the anomaly. The order parameter is the mass operator 
$M(x)=\overline{\psi}_a(x)\psi_a(x)$, which acquires a nonzero vacuum expectation value, 
giving rise to the chiral condensate which in the large-${\cal N}_c$ limit reads~\cite{q6}
\begin{equation}
\chi_{c}=\langle \overline{\psi}\psi \rangle=
-{\cal N}_{c}(\frac{g_c^{2}{\cal N}_{c}}{12\pi})^{\frac{1}{2}}\quad .
\label{zchi2}
\end{equation} 

In this section we shall exhibit the result of the computation 
of the lattice chiral condensate 
$\chi_L$ up to the fourth order in the strong coupling expansion. 
In the staggered fermion formalism $\chi_L$ is given by the expectation value of 
the mass operator 
\begin{equation}
M=-\frac{1}{Na}\sum_{x=1}^{N}\sum_{a=1}^{{\cal N}_c}(-1)^x \psi^{\dagger}_{ax}\psi_{ax}
\label{mao}
\end{equation}
on the perturbed states.
To the fourth order in $\epsilon$, $\chi_L$ is given by
\begin{equation}
\chi_{L}=-\frac{1}{a}{\cal N}_{c}(\frac{1}{2}-8\epsilon^2 +32\epsilon^4)\ .
\label{chitho}
\end{equation}

\section{Concluding Remarks}

We shall now compare the strong coupling results with the continuum theory. 
We have to extrapolate the strong coupling series, derived under the assumption 
that the parameter $\epsilon^2=t^2/g^4a^4\ll 1$, to the region in 
which $\epsilon^2\gg 1$; this region corresponds to the 
continuum theory since when $a\longrightarrow 0$, 
$\epsilon^2\longrightarrow \infty$ for a given $g$. 
For this purpose it is customary to make use of Pad\'e approximants. 

We first compute the lattice light velocity by equating the lattice chiral 
condensate Eq.(\ref{chitho}) to its continuum counterpart Eq.(\ref{zchi2});
we get $t=0.9757$ which lies $2.4\%$ below the exact answer $t=1$. 
Applying the same procedure to the mass of the pseudoscalar excitation 
the extrapolated value of $m_P$ is ${m_{P}}/{g}=0.6655$ 
which agrees within 16\% with the result obtained in~\cite{q7} in the continuum
for ${\cal N}_c=3$ and with the lattice numerical calculations of Ref.~\cite{q8}.

The strong coupling perturbation expansion even for the 't Hooft model provides, not only
accurate numerical results for the spectrum and chiral condensate, but also
leads to a simple and correct understanding of confinement and to an intuitive picture of
the vacuum of gauge theories.

\end{document}